\documentclass[10pt,a4paper]{article}               
\usepackage[utf8x]{inputenc}  
\usepackage[T1]{fontenc}  
\usepackage[english]{babel}  
\usepackage{plain}
\usepackage{hyperref}
\usepackage{amssymb}
\usepackage{amsmath}
\usepackage[margin=1in]{geometry}
\usepackage{authblk}
\usepackage{graphicx}
\date{} 

\title{Spin Quantum Entanglement in a General Curved Static Space-Time }
\author[1]{A. Mohadi\thanks{Corresponding author: \texttt{aicha.mohadi@univ\-msila.dz}}}
\author[2]{N. Mebarki\thanks{nnmebarki@yahoo.com}}
\author[3]{M.Boussahel\thanks{ mounir.boussahel@univ\-msila.dz}}
\affil[1,2]{Laboratoire de Physique Mathematique et Subatomique\\
Mentouri University, Constantine, Algeria}
\affil[3]{Laboratoire de physique et chimie des materiaux\\ Mohamed Boudiaf University, M'sila}
\begin{document}
\maketitle
\begin{abstract}
A general formalism of the spin quantum entanglement in a curved
space-time represented. As examples Kerr and non commutative Reissner-
Nordström models are considered. The behaviors of the concurrence and
entanglement entropy as a function of the various parameters are
also discussed.
\end{abstract}
\section{Introduction}
During the last decade, a great interest has been devoted to quantum
entanglement and information theory\cite{1}\cite{2}\cite{3}. The spin quantum entanglement of  a
bipartite system plays an important rule in most of the physical system such
as condensed matter. Recently, the effect of the relativistic motion on the quantum spin
states entanglement correlation has been the focus of many people\cite{4}\cite{5}\cite{6}. More, interesting was the study of the quantum entanglement of a spin system in a non inretial
frames and a curved space-time\cite{7}\cite{8}\cite{20}\cite{15} .

In this paper, we discuss the effect of a certain static gravitational field
( near to black hole) on the quantum spin entanglement (QSE) of a bipartite
system.
In section 2, we present the general mathematical formalism. In section 3 we consider the Kerr space-time.
 In section 4, the non commutative Reissner-Nordstrom space-time is considered and in section 5 we draw our conclution.
\section{Mathematical formalism}
In order to study the spin of particle in a curved space-time one has to use an inertial local frame at each point. This can be
done at the tangent at a point of curved space-time using the vierbien (or tetrad) $e_{a}^{\mu}$ ($\mu$ (resp. $a$) is a curved (resp.flat) index)  
defined by:
\begin{equation}
g_{\mu \nu} e_{a}^{\mu} e_{b}^{\nu} = \eta_{ab}
\end{equation}
where $g_{\mu \nu }$ and $\eta _{ab}$ are the metric of the curved and
Minkowski space-time respectively. Let us introduce a one fermionic particle
state $\left\vert P,\sigma \right\rangle $ with a 4-momentum $P^{\mu }$ and
spin $\sigma (=\uparrow ,\downarrow )$ at some point of the space-time. If
we move from one point to another, this state becomes (in a local frame)\cite{9}\cite{10}
\begin{equation}
\sum_{\sigma ^{^{\prime }}}D_{\sigma ^{^{\prime }}\sigma }(\Lambda
,p)\left\vert \Lambda p,\sigma ^{^{\prime }}\right\rangle
\end{equation}
where $\Lambda $ is the Lorentz transformation matrix and $D_{\sigma
^{^{\prime }}\sigma }$ the wigner rotation matrix elements \cite{11}.

Let us consider a system of two non-interacting spin $\frac{1}{2}$ particles
where its center of mass system can be described by an initial wave packet $%
\left\vert \Psi ^{i}\right\rangle $ given in a local frame by:

\begin{equation}
 \left\vert \Psi ^{i}\right\rangle =\underset{\sigma _{1}\sigma _{2}}{\sum }%
\int dp_{1}dp_{2}g_{\sigma _{1}\sigma _{2}}(p_{1},p_{2})\left\vert
p_{1},\sigma _{1};p_{2},\sigma _{2}\right\rangle 
\end{equation}
with the normalisation condition:
\begin{equation}
\underset{\sigma _{1}\sigma _{2}}{\sum }\int dp_{1}dp_{2}\left\vert
g_{\sigma _{1}\sigma _{2}}(p_{1},p_{2})\right\vert ^{2}=1
\end{equation}%
Here $p_{1}$and $p_{2}$ are the four-momentum of the particles 1 and 2
respectively. Now, it is easy to show that when the system reaches another
point of the inertial local frame, the wave packet becomes $\left\vert \Psi
^{f}\right\rangle $ such that :
\begin{equation}
\left\vert \Psi ^{f}\right\rangle =\underset{\sigma _{1}\sigma _{2}\sigma
_{1}^{^{\prime }}\sigma _{2}^{^{\prime }}}{\sum }\int dp_{1}dp_{2}\sqrt{%
\frac{(\Lambda p_{1})^{0}(\Lambda p_{2})^{0}}{p_{1}^{0}p_{2}^{0}}}g_{\sigma
_{1}\sigma _{2}}(p_{1},p_{2})\times D_{\sigma _{1}^{^{\prime }}\sigma
_{1}}(\Lambda ,p_{1})D_{\sigma _{2}^{^{\prime }}\sigma _{2}}(\Lambda
,p_{2})\left\vert \Lambda p_{1},\sigma _{1}^{^{\prime }};\Lambda
p_{2},\sigma _{2}^{^{\prime }}\right\rangle
\end{equation}%
while the change in the vierbien $\delta e_{\mu }^{a}(x)$ is given by:
\begin{equation}
\delta e_{\mu }^{a}(x)=u^{\nu }(x)d\tau \nabla _{\nu }e_{\mu
}^{a}(x)=-u^{\nu }(x)\omega _{vb}^{a}e_{\mu }^{b}(x)d\tau =\chi
_{b}^{a}(x)e_{\mu }^{b}(x)d\tau
\end{equation}%
where
\begin{equation}
\omega _{vb}^{a}=-e_{\mu }^{a}(x)\nabla _{v}e_{b}^{\mu }(x)
\end{equation}%
with 
\begin{equation}
\chi _{b}^{a}(x)=-u^{\nu }(x)\omega _{vb}^{a}  \label{C}
\end{equation}%
Here $\omega _{vb}^{a}$ and $\chi _{b}^{a}(x)$ are the spin connection and
the change of the spin connection along the direction of the 4-vector
velocity $u^{\nu }(x)$ and $\nabla _{v}$ stands for the covariant
derivative. It is worth to mention also that during this displacement, the
change in the momentum $\delta p^{\mu }(x)$ is:
\begin{equation}
\delta p^{\mu }(x)=u^{\nu }(x)d\tau \nabla _{\nu }p^{\mu }(x)=ma^{\mu
}(x)d\tau
\end{equation}%
where $\tau $ is the proper time and $a^{\mu }(x)$ the 4-verctor
acceleration given by:

\begin{equation}
a^{\mu }(x)=u^{\nu }(x)\bigtriangledown _{\nu }u^{\mu }(x)
\end{equation}%
Straightforward simplifications lead to:
\begin{equation}
\delta p^{a}(x)=\lambda _{b}^{a}(x)p^{b}(x)d\tau 
\end{equation}%
where the infinitisimal Lorentz transformation matrix elemnts $\lambda
_{a}^{b}(x)$ have the form:

\begin{equation}
\lambda _{b}^{a}(x)=-\frac{1}{mc^{2}}[a^{a}(x)p_{b}(x)-p^{a}(x)a_{b}(x)]+%
\chi _{b}^{a}(x)  \label{D}
\end{equation}%
and
\begin{equation}
a^{\mu }(x)=u^{\nu }[\partial _{\nu }u^{\mu }+\Gamma _{\nu \lambda }^{\mu
}u^{\lambda }]
\end{equation}%
where $\Gamma _{\nu \lambda }^{\mu }$ is the affine connection.

Now; let us consider a general static universe where the mertic $ds^{2}$ has
the form:

\begin{equation}
ds^{2}=F(r)dt^{2}+G(r)dr^{2}+H(r,\theta )d\theta ^{2}+I(r,\theta )d\varphi
^{2}  \label{A}
\end{equation}%
Using the spherical coordinate ($r,\theta ,\varphi $), and choosing the
tetrad components:
\begin{equation}
e_{0}^{t}=\frac{1}{\sqrt{F(r)}}\text{ \ \ \ }e_{1}^{r}=\frac{1}{\sqrt{G(r)}}%
\text{\ \ \ }e_{2}^{\theta }=\frac{1}{\sqrt{H(r,\theta )}}\text{ \ \ \ }%
e_{3}^{\varphi }=\frac{1}{\sqrt{I(r,\theta )}}
\end{equation}%
Thus, the non vanishing spin connection elements are :
\begin{eqnarray}
\omega _{t1}^{0} &=&\frac{1}{2}\frac{F^{\prime }}{\sqrt{GF}}\ \ \ \ \ \omega
_{\varphi 3}^{0}=\frac{\overset{\bullet }{I}}{2\sqrt{IF}}\ \ \ \ \ \omega
_{\theta 2}^{1}=-\frac{1}{2\sqrt{GH}}H^{\prime }\ \  \\
\ \ \  \omega _{\varphi 3}^{1} &=&-\frac{1}{2}\frac{I^{\prime }}{\sqrt{GI}}%
\ \ \ \ \ \omega _{\varphi 3}^{2}=-\frac{1}{2\sqrt{HI}}\partial _{\theta }I
\notag
\end{eqnarray}%
where $\overset{\bullet }{I}=\frac{\partial I}{\partial t}$ and $I^{\prime }=%
\frac{\partial I}{\partial r}$. Furthermore the non vanishing components $%
u^{\nu },\chi _{b}^{a}$ and $\lambda _{a}^{b}$, for a circular motion and
constant angular velocity $\frac{d\varphi }{dt}$ on the equatorial plane
where $\theta =\frac{\pi }{2}$ are given by:

\begin{equation}
u^{t}=\frac{\gamma c}{\sqrt{F}}\ \ \ \ \ \ \ \ \ \ \ \ \ \ \ u^{\varphi }=%
\frac{1}{\sqrt{I}}\gamma r\frac{d\varphi }{dt}
\end{equation}

\begin{eqnarray}
\chi _{1}^{0} &=&-u^{t}\ \omega _{t1}^{0}\ \  \ \chi _{3}^{0}=-u^{\varphi
}\omega _{\varphi 1}^{0}\ \ \  \ \ \chi _{2}^{1}=-u^{\theta }\omega
_{\theta 2}^{1}\ \  \\\
\ \ \ \ \ \chi _{3}^{1} &=&-u^{\varphi }\omega _{\varphi 3}^{1}\ \ \ \ \ \
\ \ \chi _{3}^{2}=-u^{\varphi }\omega _{\varphi 3}^{2}  \notag
\end{eqnarray}%
and:
\begin{eqnarray}
\lambda _{1}^{0} &=&\frac{1}{mc^{2}}[p^{0}a_{1}]+\chi _{1}^{0}\ \ \ \ \ \ \
\lambda _{3}^{1}=-\frac{1}{mc^{2}}[a^{1}p_{3}]+\chi _{3}^{1}\ \ \ \  \\
\ \ \ \lambda _{3}^{2} &=&\chi _{3}^{2}\ \ \ \ \ \ \ \ \ \ \ \ \ \ \ \ \ \ \
\ \ \ \ \ \ \ \ \ \ \lambda _{2}^{0}=\chi _{2}^{0}  \notag
\end{eqnarray}
It is important to mention that the two non vanishing components of the
4-vector velocity $u^{t}$ and $u^{\varphi }$ can be rewritten as:
\begin{equation}
u^{t}=\frac{c\cosh \xi }{\sqrt{f(r)}}\ \ \ \ \ \ \ \ \ \ \ \ \ \ \
u^{\varphi }=\frac{c\sinh \xi }{\sqrt{I(r,\theta )}}
\end{equation}%
where $\xi $ is the rapidity in the local inertial frame such that $\frac{v%
}{c}=\tanh \xi $ ($\frac{v}{c}=\sqrt{\frac{\gamma ^{2}-1}{\gamma ^{2}}}%
,\gamma $ is the Lorentz factor)
To quantify the spin entanglement of the two particles system, we use the
Wootters concurrence\cite{12,13}, for the mixed state $\left\vert
p_{1},\uparrow ,p_{2},\downarrow \right\rangle $ defined by:

\begin{equation}
C(\rho )=\max \{0,\sqrt{\lambda _{1}}-\sqrt{\lambda _{2}}-\sqrt{\lambda _{3}}%
-\sqrt{\lambda _{4}}\}  \label{E}
\end{equation}%
where $\rho $ is the state density matrix and $\sqrt{\lambda _{i}}$ are the
eigenvalues of $\rho \overset{\thicksim }{\rho }$ \ where $\overset{%
\thicksim }{\rho }=(\sigma _{y}\otimes \sigma _{y})\rho ^{\ast }(\sigma
_{y}\otimes \sigma _{y})$ with $\sigma _{y}$ is the Pauli matrix. If $%
\lambda _{i}$ are positive real numbers, the entanglement can be quantified
by the entanglement entropy $E(\varrho )$\ defined as\cite{6}:

\begin{equation}
E(\varrho )=h(\frac{1+\sqrt{(1-C^{2}(\rho ))}}{2})
\end{equation}%
where%
\begin{equation}
h(x)=-x\log _{2}x-(1-x)\log _{2}(1-x)
\end{equation}%
In the case of a curved static space-time and spin singlet state, eq(\ref{E}%
)can be shown to have the following expression\cite{8}:

\begin{equation}
C(\varrho ^{f})=\left\langle \cos \Theta \right\rangle ^{2}+\left\langle
\sin \Theta \right\rangle ^{2}  \label{F}
\end{equation}%
where:
\begin{equation}
\langle \cos x \rangle =\int\left\vert f(p)\right\vert^{2}\cos x\, \mathrm{d}p \label{f}
\end{equation}%
and $\Theta $ is a shorthandnotation for $\tau \Theta _{3}^{1}$ ($\Theta
_{3}^{1}$ is the only non vanishing components of $\Theta _{k}^{i}$) with:

\ \ \ 
\begin{equation}
\ \ \ \ \ \Theta _{k}^{i}=\lambda _{k}^{i}+\frac{\lambda
_{0}^{i}p_{k}-\lambda _{k0}p^{i}}{p^{0}+mc^{2}}
\end{equation}%
For the general static metric of eq(\ref{A}), $\Theta $ can be rewritten as:
\begin{equation}
\Theta =\Theta _{3}^{1}\tau =-\frac{\alpha }{2G}\{[AD^{2}+B(D^{2}-1)-I^{%
\prime }\sqrt{\frac{G}{HI}}]+\frac{D}{C}p[AD^{2}+B(D^{2}-1)+A\sqrt{G}]\}
\end{equation}%
where $A=\frac{F^{\prime }}{F};B=\frac{I^{\prime }}{H},D=\sqrt{1+q^{2}},C=1+%
\sqrt{1+p^{2}}$ and $\alpha =\frac{\tau }{r_{s}}$

\section{ Kerr space-time}

As a first application we consider the Kerr metric. In Boyer-Lindquits coordinates has the following expression%
\cite{16}:

\begin{equation}
ds^{2}=\frac{\Delta }{\rho ^{2}}d\widehat{t}^{2}-\frac{\rho ^{2}}{\Delta }%
g(r)dr^{2}-\rho ^{2}d\theta ^{2}-\frac{\sin ^{2}\theta }{\rho ^{2}}d\widehat{%
\varphi }^{2}
\end{equation}
where 
\begin{eqnarray}
\rho &=&r^{2}+a^{2}\cos ^{2}\theta \text{ \ \ \ \ \ \ \ }\Delta
=r^{2}+r_{s}r+a^{2}+r_{Q}^{2}\text{ \ \ } \\
a &=&\frac{J}{Mc}\text{ \ \ \ \ \ \ \ \ \ \ \ \ \ \ \ \ \ }r_{s}=\frac{2GM}{%
c^{2}}\text{ \ \ \ \ \ \ \ \ \ \ \ \ \ \ \ \ \ \ }r_{Q}^{2}=\frac{Q^{2}G}{%
4\pi \zeta _{0}c^{4}}  \notag
\end{eqnarray}
and 

\begin{eqnarray*}
d\widehat{t}^{2} &=&(dt-a\sin ^{2}\theta d\phi )^{2} \\
d\widehat{\varphi }^{2} &=&((r^{2}+a^{2})d\phi -adt)^{2}
\end{eqnarray*}

\begin{figure}[htb!]
\begin{center}

  \includegraphics[width=12cm,height=5cm]{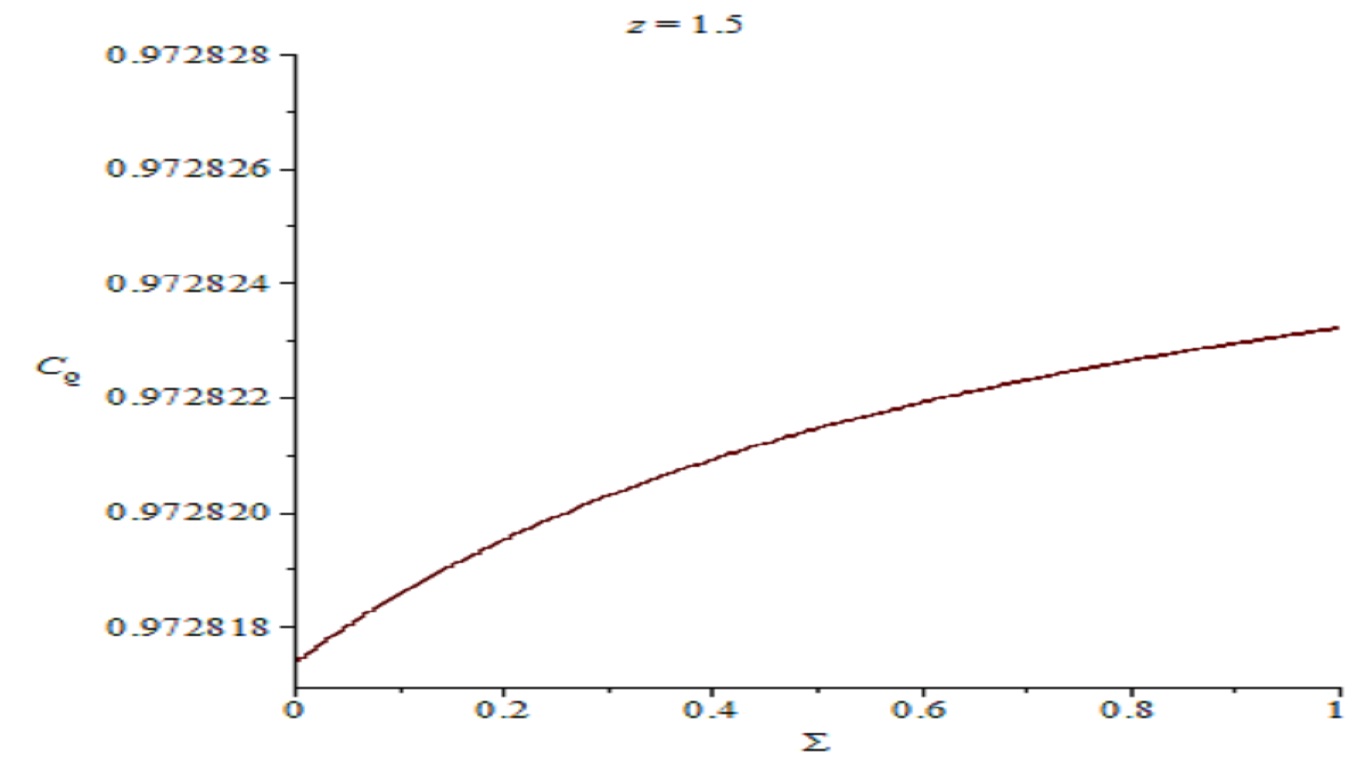}
  \caption{Variation of the concurrence as a function
of $\Sigma $ with fixed $\protect\alpha =1,z=1.5$ and $q=0.1$.}
  \label{fig:Q1KPHT00}
\end{center}
\end{figure}

\begin{figure}[htb!]
\begin{center}
  \includegraphics[width=12cm,height=5cm]{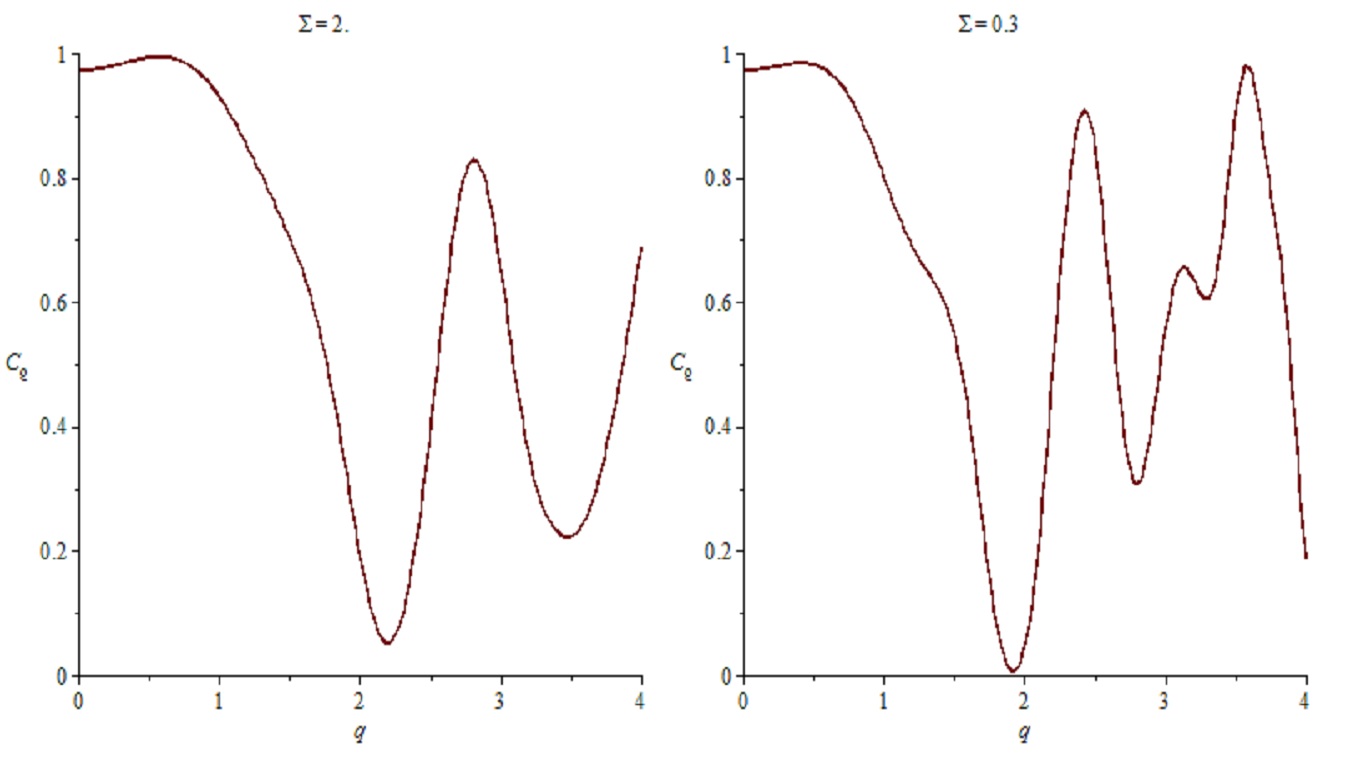}
 \caption{The concurrence as a function of $q$ for fixed $\protect\alpha =2,z=1.5$.}
  \label{fig:Q1LC7G00}
\end{center}
\end{figure}

\begin{figure}[htb!]
\begin{center}
  \includegraphics[width=12cm,height=5cm]{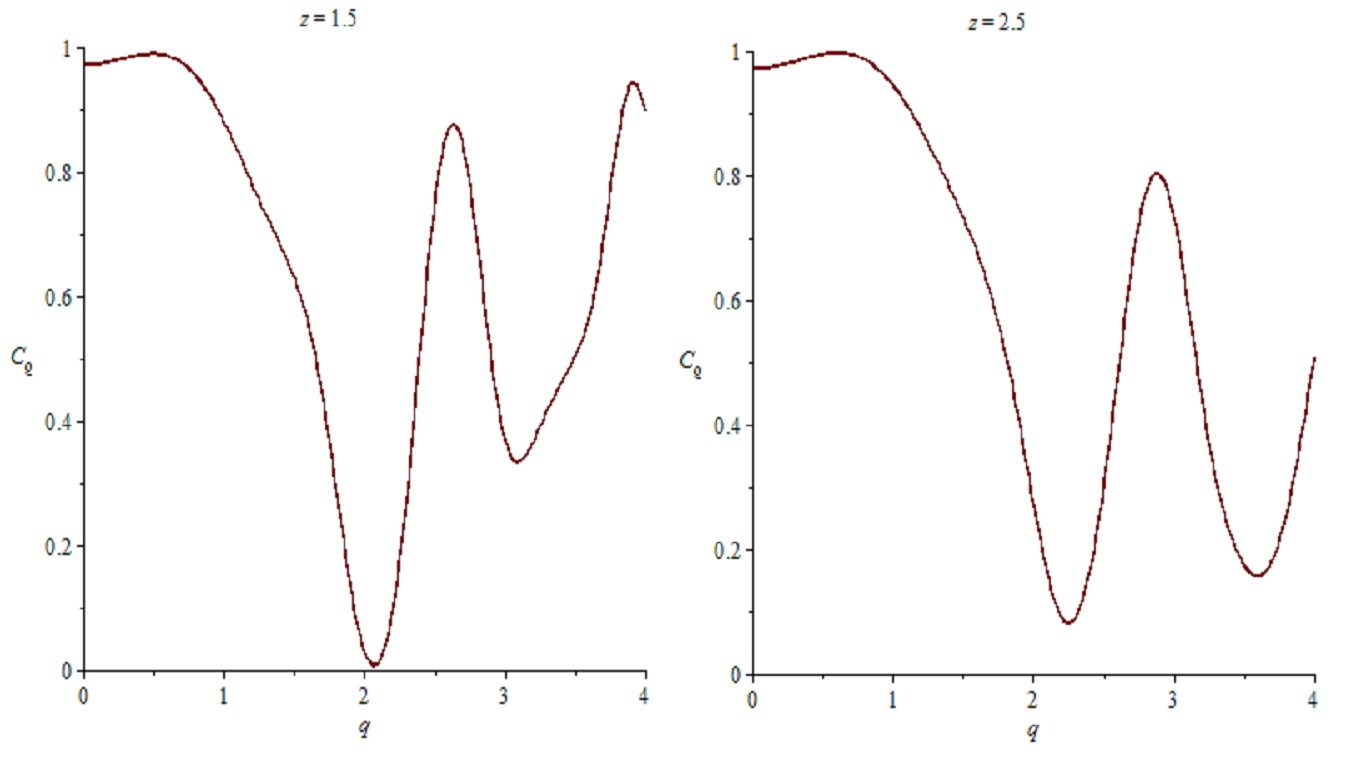}
 \caption{The concurrence as a function of $q$ for fixed $\protect\alpha =1,\Sigma =1$.}
  \label{fig:Q1KSCD0B}
\end{center}
\end{figure}

\begin{figure}[htb!]
\begin{center}
  \includegraphics[width=12cm,height=5cm]{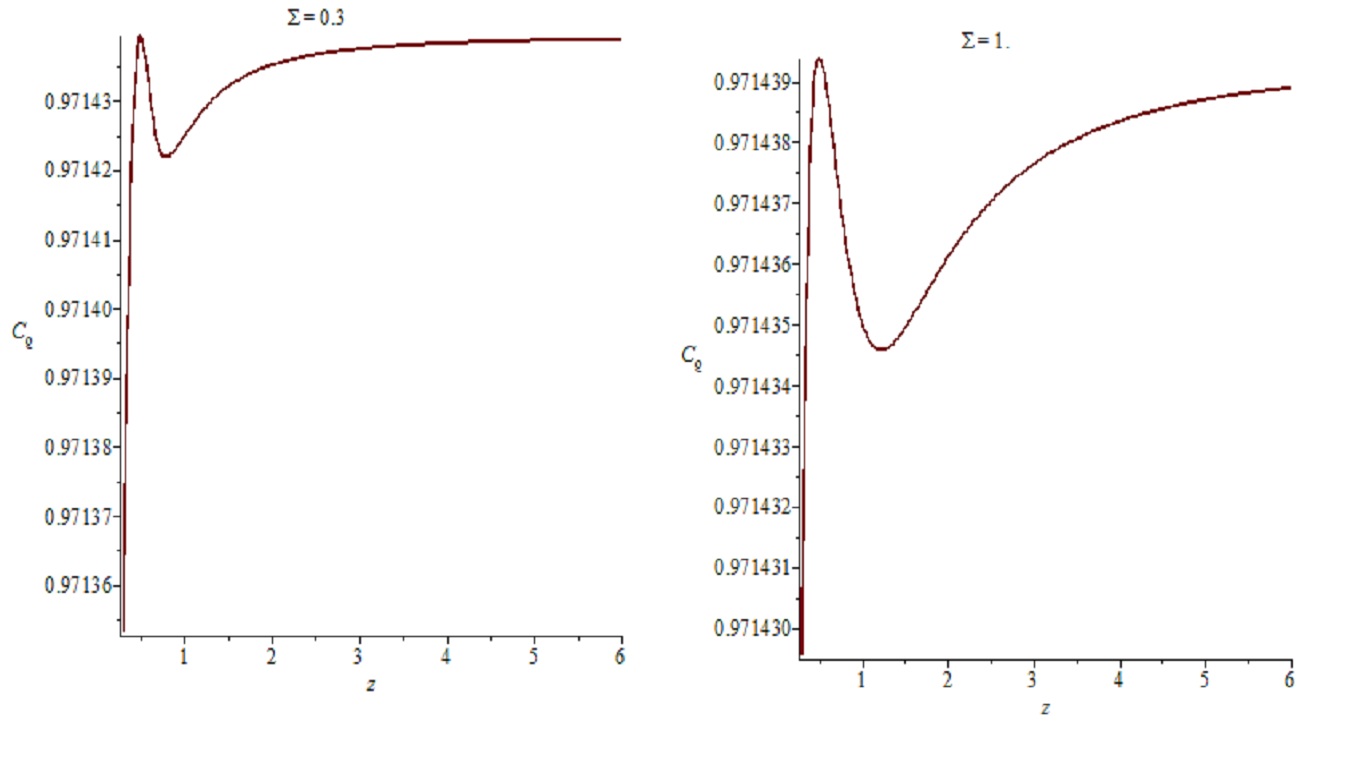}
 \caption{ the concurrence as a function of $z$ for fixed $q=0.1,\protect\alpha =1$.}
  \label{fig:Q1MDWL00}
\end{center}
\end{figure}

\begin{figure}[htb!]
\begin{center}
  \includegraphics[width=12cm,height=5cm]{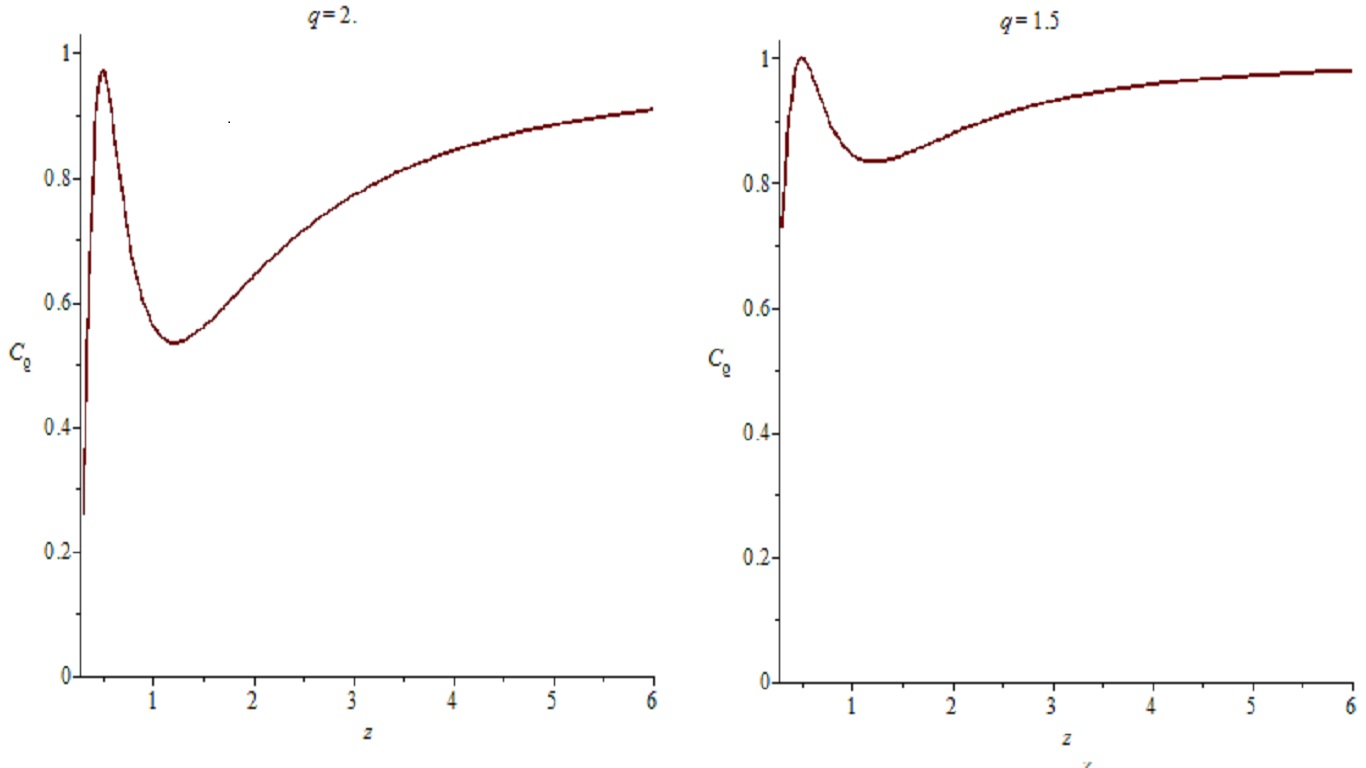}
 \caption{ the concurrence as a function of $z$ for fixed $\Sigma=1,\protect\alpha =1$.}
  \label{fig:Q1MHC002}
\end{center}
\end{figure}

\begin{figure}[htb!]
\begin{center}
  \includegraphics[width=6cm,height=5cm]{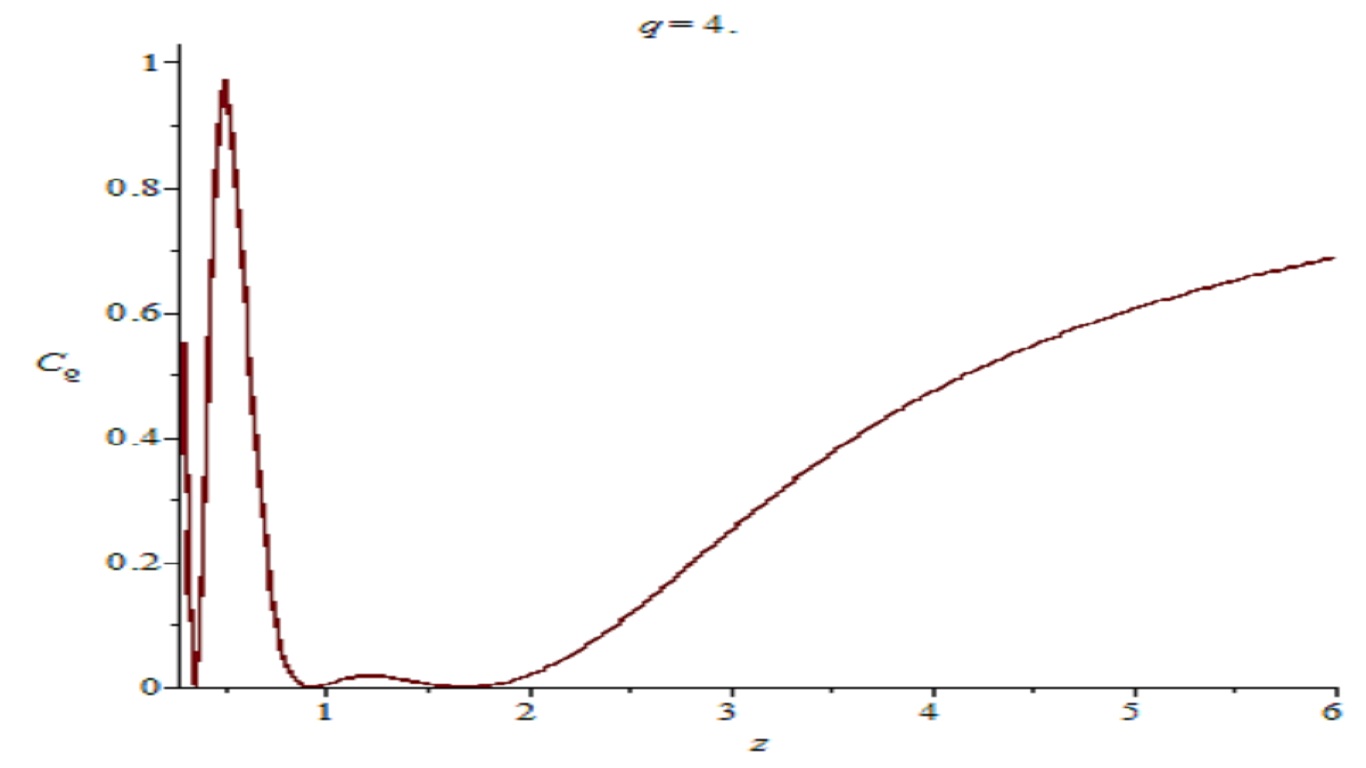}
 \caption{ the concurrence as a function of $z$ for fixed $q=0.1,\protect\alpha =1$.}
  \label{fig:Q1MME103}
\end{center}
\end{figure}

Here $Mc,J$ and $Q$ are the mass, angular momentum and charge $Q$ of the
blak hole, $G,c,\zeta _{0}$ are the Newton gravitational constant, velocity
of light, vacuum permittivity respectively. In what follows we deal with a
non charged black hole where $Q=0.$ In this case, we can show that the non
vanishing components are:

\begin{equation}
e_{t}^{0}=\sqrt{\frac{\Delta }{r^{2}}}\text{ \ \ \ \ \ \ }e_{r}^{1}=\sqrt{%
\frac{r^{2}}{\Delta }}\text{ \ \ \ \ }e_{\theta }^{2}=\sqrt{r^{2}}\text{ \ \
\ \ \ \ }e_{\varphi }^{3}=\sqrt{\frac{1}{r^{2}}}
\end{equation}

\begin{equation}
u^{t}=r\frac{\gamma c}{\sqrt{\Delta }}\ \ \ \ \ \ \ \ \ \ \ \ \ \ \ \ \ \ \
\ \ \ \ \ \ \ \ \ \ \ \ \ \ \ \ \ \ \ \ \ \ \ \ u^{\varphi }=r\gamma v
\end{equation}

\begin{eqnarray}
\Gamma _{rr}^{r} &=&-\frac{r^{2}r_{s}-2ra^{2}}{2\Delta r^{2}}\ \ \ \ \ \ \ \
\ \ \ \Gamma _{tt}^{r}=\frac{\Delta (rr_{s}-2a^{2})}{2r^{5}}\ \ \ \ \ \ \ \
\ \Gamma _{\varphi \varphi }^{r}=\frac{\Delta }{r^{5}} \\
\Gamma _{\theta \theta }^{r} &=&-\frac{\Delta }{r}\ \ \ \ \ \ \ \ \ \ \ \ \
\ \ \ \ \ \ \ \ \ \ \ \ \ \Gamma _{rt}^{t}=\frac{r^{2}r_{s}-2ra^{2}}{2\Delta
r^{2}}\text{ \ \ \ \ \ \ \ \ \ }\ \Gamma _{r\theta }^{\theta }=\frac{1}{r} 
\notag
\end{eqnarray}

\begin{equation}
\omega _{t1}^{0}=\frac{rr_{s}-2a^{2}}{2r^{3}}\ \ \ \ \ \ \ \ \ \ \ \ \ \ \ \
\ \ \ \ \ \ \omega _{\theta 2}^{1}=\sqrt{\Delta }\ \ \ \ \ \ \ \ \ \ \ \ \ \
\ \ \ \ \ \ \omega _{\varphi 3}^{1}=\frac{\sqrt{\Delta }}{r^{3}}
\end{equation}

\begin{equation}
\chi _{1}^{0}=-\frac{\gamma c}{\sqrt{\Delta }}\frac{rr_{s}-2a^{2}}{2r^{2}}\
\ \ \ \ \ \ \ \ \ \ \ \ \ \ \ \ \ \ \ \ \ \ \ \ \ \ \ \ \chi
_{3}^{1}=-\gamma v\frac{\sqrt{\Delta }}{r^{2}}
\end{equation}

\begin{equation}
a^{r}=\frac{\gamma ^{2}c^{2}(rr_{s}-2a^{2})}{2r^{3}}+\gamma ^{2}v^{2}\frac{%
\Delta }{r^{3}}
\end{equation}

\begin{eqnarray}
\ \lambda _{1}^{0}\ &=&\frac{1}{mc^{2}}p^{0}\gamma ^{2}v^{2}[\frac{(r_{s}-2r)%
}{2r^{2}}]\sqrt{\frac{r^{2}}{\Delta }}\ \ \ \ \ \ \ \  \\
\ \ \ \ \ \ \ \ \ \ \lambda _{3}^{1}\ &=&-\frac{1}{mc^{2}}[\frac{\gamma
^{2}c^{2}(-r_{s}+2r)}{2r^{2}}]\sqrt{\frac{r^{2}}{\Delta }}q^{3}  \notag
\end{eqnarray}%
Now, if we set $q=\frac{q^{3}}{mc},p=\frac{p^{3}}{mc},z=\frac{r}{r_{s}}%
,\Sigma =\frac{a}{r_{s}},\alpha =\frac{\tau c}{r_{s}},$ one has$:$

\begin{equation}
\Theta =\alpha \sqrt{\frac{z^{2}}{z^{2}-z+\Sigma }}[\frac{(1-2z)}{2z^{2}}]q%
\sqrt{(q^{2}+1)}[\sqrt{(q^{2}+1)}-\frac{qp}{\sqrt{(p^{2}+1)}+1}]
\end{equation}%
Figure \ref{fig:Q1KPHT00} displays the variation of the concurrence as a function of the
dimensionless parameter $\Sigma \in \lbrack 0.1]$ and fixed $\alpha \approx
1,z=1.5$ and $q\approx 0.1$, the concurrence is an increasing function. This
is due to the fact that the gravitational potential $g_{00}$ decreases as $J$
(or $\rho $) increases and thus information (or concurrence) increases until
a saturated bound of the maximal entanglement ( $C(\varrho ^{f})\sim 1$).
 Figure \ref{fig:Q1LC7G00} shows the concurrence variation as a function of $q$ for fixed values
of $\alpha =2,z$ and $\Sigma $, at smaller value of $q$ ($q\rightarrow 0$),
the entangled is max and if $q^{\nearrow }$ the center of the wave packet
travels more on the circular trajectory and therefore one has more
decoherence (less entanglement ) and consequently the concurrence decreases
for example if $q=1$ \ ,$C(\varrho ^{f})=0.7$ and if $q=1.6$ \ $C(\varrho
^{f})=0.4$, the oscillator periodic behavior can be explained (as it was
pointed out in ref\cite{15}.) by the fact that when $q$ increases, the
exponential in the integral that present in the expression of the
concurrence approaches unity, so the cosine and sine terms behavior
dominates. It is worth to mention that this behavior (minima and maxima) changes if
the other parameters such as $\Sigma $ and $z$ changes. Figure \ref{fig:Q1LC7G00} shows that if 
$\Sigma $ decreases to $0.3$ the number and shape of picks change and they
become more pronounced, similar behavior is shown in Figure \ref{fig:Q1KSCD0B} if $z$ changes.
\\Figure \ref{fig:Q1MDWL00} and \ref{fig:Q1MHC002} display the variation of the concurrence as a function of $%
z$ (or circular motion radius) for fixed values of $q,\Sigma ,\alpha =1,$
notice that for smaller valeus of $r$ ($r\rightarrow 0$ near black hole
singularity) where the gravitational field is infinite, the entanglement is
minimal ($C(\varrho ^{f})\sim 0$). if we go far from the singularity ($z$
increases) the gravitational field decreases and therefore the information
increases and thus the concrrence ($C(\varrho ^{f})\sim 1$). The shape and
number of picks and minima depend strongly on the values of the parameters $%
q $ and $\Sigma $. Figures \ref{fig:Q1MDWL00}, \ref{fig:Q1MHC002} and \ref{fig:Q1MME103} show the behavior of the
concurrence with variation of $\Sigma $ and $q$ respectively.
\clearpage
\section{ Reissner-Nordstr\"{o}m noncommutative space-time}

As a second example we consider the Reissner-Nordstr\"{o}m metric for a charged non rotating black hole in a
commutative space-time. It is given by \cite{17}:

\begin{equation}
ds^{2}=-(1-\frac{2M}{r}+\frac{Q2}{r^{2}})c^{2}dt^{2}+\frac{dr^{2}}{(1-\frac{%
2M}{r}+\frac{Q2}{r^{2}})}+r^{2}d\theta ^{2}+r^{2}\sin ^{2}\theta d\varphi
^{2}
\end{equation}

with $M$ and $Q$ are mass and charge. Following ref\cite{14}, the Seiberg
Witten vierbein $\widehat{e}_{\mu }^{a}$ in a noncommutative gauge gravity
is given by:

\begin{equation}
\widehat{e}_{\mu }^{a}=e_{\mu }^{a}(x)-i\widetilde{\eta }^{\nu \rho }e_{\mu
v\rho }^{a}(x)+\widetilde{\eta }^{\nu \rho }\widetilde{\eta }^{\lambda \tau
}e_{\mu v\rho \lambda \tau }^{a}(x)+O(\widetilde{\eta }^{3})
\end{equation}%
where

\begin{equation}
e_{\mu v\rho }^{a}=\frac{1}{4}[\omega _{v}^{ac}\partial _{\rho }e_{\mu
}^{d}+(\partial _{\rho }^{a}\omega _{\mu }^{ac}+R_{\rho \mu
}^{ac})e_{v}^{d}]\eta _{cd}
\end{equation}%
and

\begin{eqnarray}
e_{\mu v\rho \lambda \tau }^{a} &=&\frac{1}{32}[2\{R_{\tau v},R_{\mu \rho
}\}^{ab}e_{\lambda }^{c}-\omega _{\lambda }^{ab}(D_{\rho }R_{\tau \mu
}^{cd}+\partial _{\rho }R_{\tau \mu }^{cd})e_{v}^{m}\eta _{dm} \\
&&-\{\omega _{\nu },(D_{\rho }R_{\tau \mu }+\partial _{\rho }R_{\tau \mu
})\}^{cd}e_{\lambda }^{c}-\partial _{\tau }\{\omega _{\nu },(\partial _{\rho
}\omega _{\mu }+R_{\rho \mu })\}^{ab}e_{\lambda }^{c}  \notag \\
&&-\omega _{\lambda }^{ab}\partial _{\tau }(\omega _{\nu }^{cd}\partial
_{\rho }e_{\mu }^{m}+(\partial _{\rho }\omega _{\mu }^{cd}+R_{\rho \mu
}^{cd})e_{\mu }^{m})\eta _{dm}+2\partial _{v}\omega _{\lambda }^{ab}\partial
_{\rho }\partial _{\tau }e_{\mu }^{c}  \notag \\
&&-2\partial _{\rho }(\partial _{\tau }\omega _{\mu }^{ab}+R_{\tau \mu
}^{cd})\partial _{v}e_{\lambda }^{c}-\{\omega _{\nu },(\partial _{\rho
}\omega _{\lambda }+R_{\rho \lambda })\}^{ab}\partial _{\tau }e_{\mu }^{c} 
\notag \\
&&-(\partial _{\tau }\omega _{\mu }^{ab}+R_{\tau \mu }^{ad})(\omega _{\nu
}^{cd}\partial _{\rho }e_{\lambda }^{m}+(\partial _{\rho }\omega _{\lambda
}^{cd}+R_{\rho \lambda }^{cd})e_{v}^{m}\eta _{dm})]\eta _{bc}  \notag
\end{eqnarray}%
where $\overset{\sim }{\eta }^{v\rho }$ is noncommutativity anti-symmetric
matrix elements defined as:

\begin{equation}
\left[ \widetilde{x}^{\mu },\widetilde{x}^{\nu }\right] =i\widetilde{\eta }%
^{\mu \nu }
\end{equation}%
and $\widetilde{x}^{\mu }$ are the noncommutative space-time coordinates
operators. Here $\omega _{\lambda }^{ab}$ (resp.$D_{\rho }$) is the
commutative spin connection (resp. covariante derivative) and $R_{\mu
v}^{ad}=e_{\alpha }^{a}e_{\beta }^{b}R_{\mu v}^{\alpha \beta }$ where $%
R_{\mu v}^{\alpha \beta }$ is the Riemann tensor. The commutative space-time
vierbein and Minkowski metric are denoted by $e_{\mu }^{a}$ and $\eta _{\mu
b}$ respectively. The noncommutative metric:

\begin{equation}
\widehat{g}_{\mu \nu }=\frac{1}{2}(\widehat{e}_{\mu }^{a}\ast \widehat{e}%
_{av}+\widehat{e}_{v}^{a}\ast \widehat{e}_{a\mu })
\end{equation}%
where "$\ast $" is the Moyal star product \cite{18}, Straightforward
calculations using the Maple 13 and setting $z=\frac{r}{r_{s}}$ and $y=\frac{%
Q^{2}}{r_{s}^{2}},\frac{\widetilde{\eta }^{2}}{r_{s}^{2}}=\lambda ,$ (in the
case $\theta =\frac{\pi }{2}$) one has:

\begin{eqnarray}
F &=&-(1-\frac{1}{z}+\frac{y}{z^{2}})-\frac{(2z^{3}-9yz^{2}-\frac{11}{4}%
z^{2}+15zy-14y^{2})\lambda }{4z^{6}} \\
G &=&\frac{1}{(1-\frac{1}{z}+\frac{y}{z^{2}})}+\frac{(-z^{3}+\frac{3}{4}%
z^{2}+3yz^{2}-3yz+2y^{2})\lambda }{4z^{2}(z^{2}-z+y)^{2}}  \notag \\
H &=&(z^{2}+\frac{(z^{4}-\frac{17}{2}z^{3}+\frac{17}{2}z^{2}+27yz^{2}-\frac{%
75}{2}zy+30y^{2})\lambda }{16z^{2}(z^{2}-z+y)})  \notag \\
I &=&(z^{2}+\frac{(-2z^{3}+8yz^{2}+z^{2}-8zy+8y^{2})\lambda }{16(z^{2}-z+y)})
\notag
\end{eqnarray}
\begin{figure}[htb!]
\begin{center}
  \includegraphics[width=6cm,height=5cm]{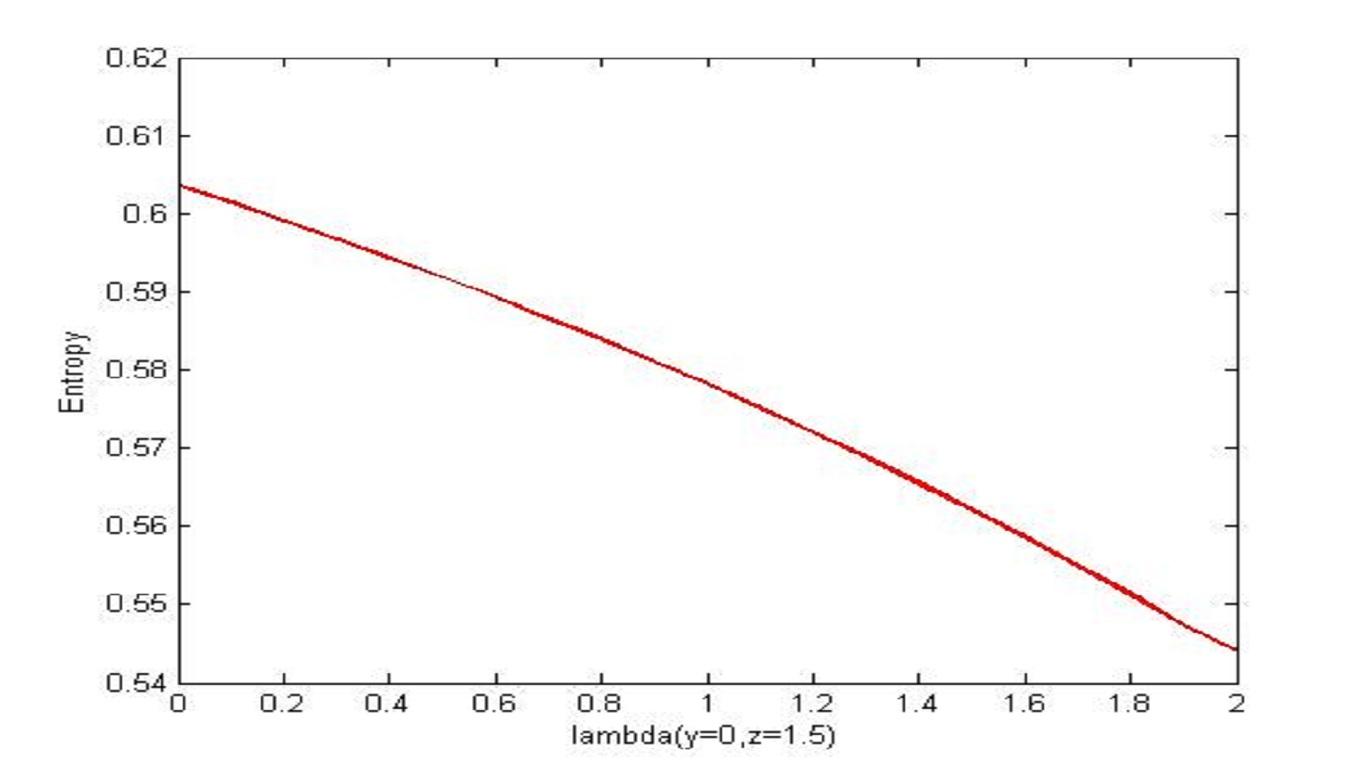}
 \caption{ $ E(\varrho) $ as a function of the $\protect\lambda $ for fixed $z=1.5,y=0,\protect\alpha %
=1,q=0.01$.}
  \label{fig:Q1W5BE00}
\end{center}
\end{figure}
 Figure \ref{fig:Q1W5BE00} displays the variation of the entanglement entropy $E(\varrho )$ as
a function of the noncommutativity parameter $\widetilde{\eta }^{2}$ for a
non charged ($Q=0$) black hole for fixed $z=1.5,y=0,\alpha =1,q=0.01$.
Notice that if $\widetilde{\eta }^{2}$ increases $E(\varrho )$ decreases.
Thus, $\widetilde{\eta }^{2}$ plays the role of a gravitational field (GF).
In fact, as it was pointed out in ref\cite{19}, the non commutativity
parameter $\widetilde{\eta }$ can be considered as like a magnetic field
contributing to the matter density $\rho $ and therefore affecting the
curvature of the space-time through its contribution to GF. Consequently if $%
\widetilde{\eta }^{2}$ increases the GF increases and the information \
decreases. Including the contribution of NC of the space-time will generate
an additional terms proportional to $\widetilde{\eta }^{2}$. In fact the
gravitational potential $\widehat{g}_{00}$ will be of the form:

\begin{equation}
\widehat{g}_{00}=\widehat{A}+\widehat{B}Q^{2}+\widetilde{\eta }^{2}(\widehat{%
D}Q^{4}+\widehat{C}Q^{2}+\widehat{F})\label{AI}
\end{equation}%
where

\begin{eqnarray}
\widehat{A} &=&-1+\frac{1}{z}\text{ \ \ \ \ \ \ \ \ \ }\widehat{B}=-\frac{1}{%
z}\frac{1}{r_{s}^{2}}\text{ \ \ \ \ \ \ \ }\widehat{D}=\frac{7}{%
z^{6}r_{s}^{6}} \\
\widehat{C} &=&(9z-15)\frac{1}{4z^{5}r_{s}^{4}}\text{ \ \ \ \ \ }\widehat{F}%
=(-2z+\frac{11}{4})\frac{1}{4z^{4}r_{s}^{4}}  \notag
\end{eqnarray}%
The behavior of the entanglement entropy $E(\varrho )$ depends strongly on
the sign of $(\widehat{D}Q^{4}+\widehat{C}Q^{2}+\widehat{F})$ having $%
\widehat{A}$ and $\widehat{B}$ negative:

1) If $Q^{2}\gg 1$ (in arbitrary unit) the term $\widehat{D}Q^{4}$
dominates, since $\widehat{D}\succ 0$, then if $\widetilde{\eta }^{2}$
increases the GF decreases leading to an increases in $E(\varrho )$ (as it
is the case of Figure \ref{fig:Q1XLRW01})\\
\begin{figure}[htb!]
\begin{center}
  \includegraphics[width=6cm,height=5cm]{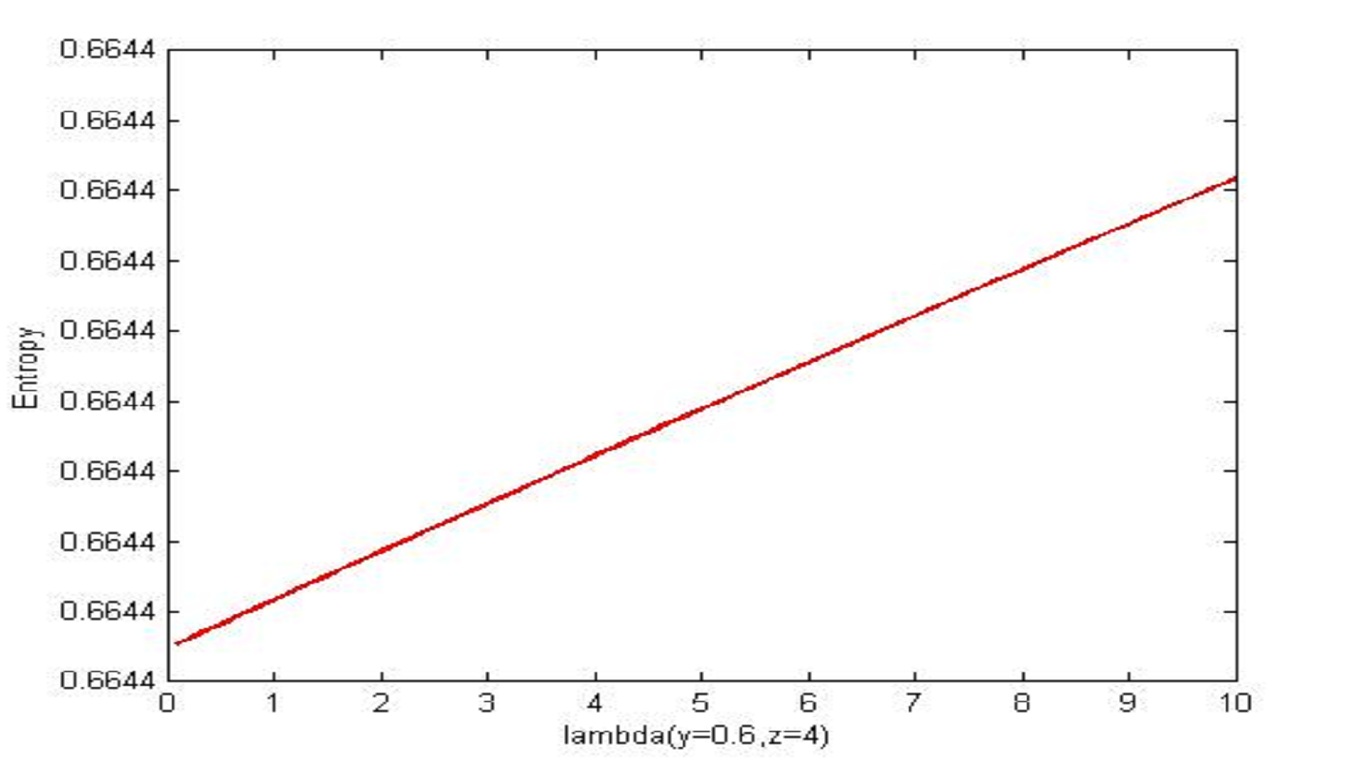}
 \caption{  $E(\protect%
\varrho )$ as a function of $\lambda$ for fixed $%
z=4,y=0.6,\protect\alpha =1,q=0.01$.}
  \label{fig:Q1XLRW01}
\end{center}
\end{figure}
2) If $Q^{2}\ll 1$, then the term $\widehat{F}$ dominates and its sign will
determine the behavior of $E(\varrho )$ as a function of $\widetilde{\eta }%
^{2}$, if $\widehat{F}\succ 0,$ then GF increases and $E(\varrho )$
decreases, we return to case in the Figure \ref{fig:Q1W5BE00}. Figure \ref{fig:Q1XLRW02} represents the variation of $E(\varrho )$ as function of $z$ for fixed $Q=0,\widetilde{\eta }=0,\alpha =1,q=0.01,$ (case of Schwarzschild
space-time in commutative space-time). Notice that we will reproduce the same behavior as in ref\cite{8}.
\begin{figure}[htb!]
\begin{center}
  \includegraphics[width=6cm,height=5cm]{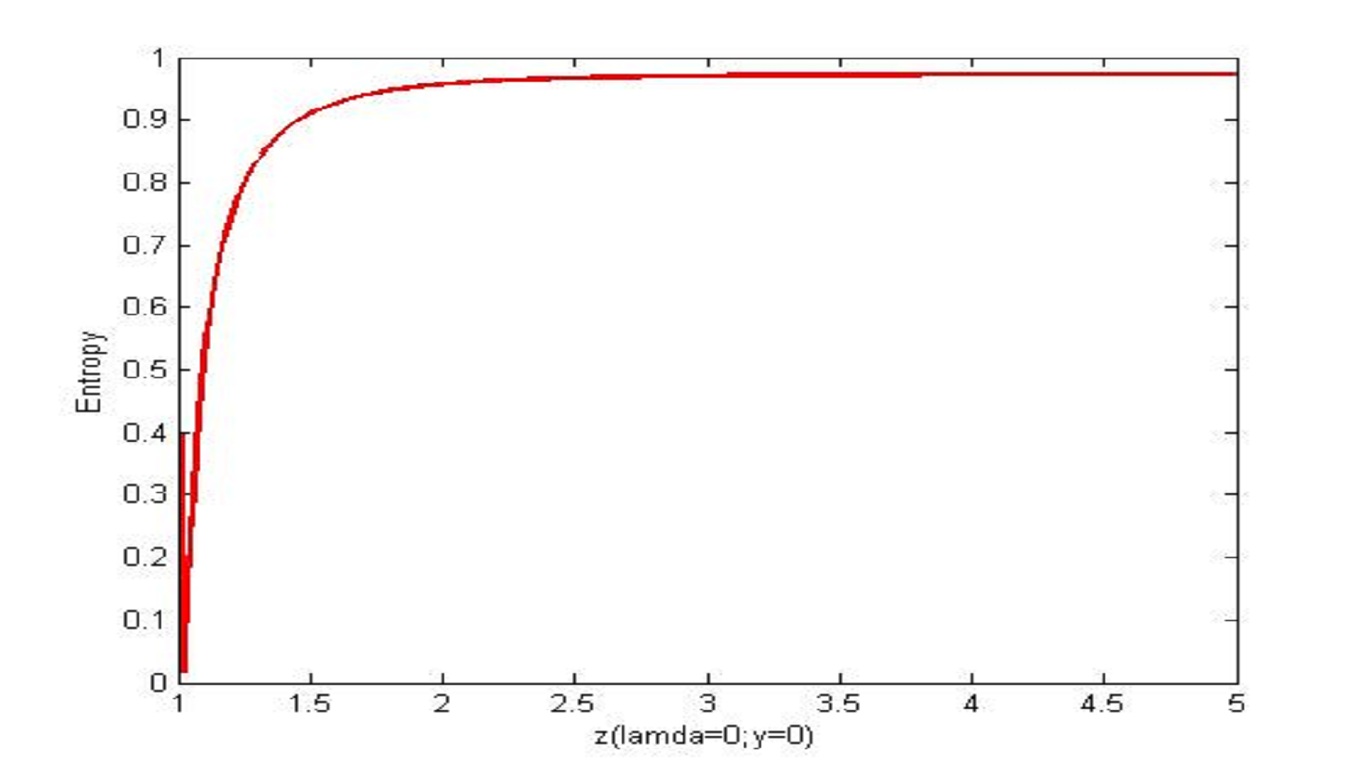}
 \caption{ the variation of $%
E(\protect\varrho )$ as a function of $z$ for fixed $\protect\lambda =0,%
\protect\alpha =1,y=0,\protect\alpha =1,q=0.01$.}
  \label{fig:Q1XLRW02}
\end{center}
\end{figure}
Figure \ref{fig:PAVXRT01} shows the variation of $E(\varrho )$ as a function of $z$ for fixed $%
Q\neq 0,\widetilde{\eta }=0$ (case of Reissner Nordstrom in commutative
space-time). Notice that the same behavior as in ref\cite{6} was obtained.
\begin{figure}[htb!]
\begin{center}
  \includegraphics[width=6cm,height=5cm]{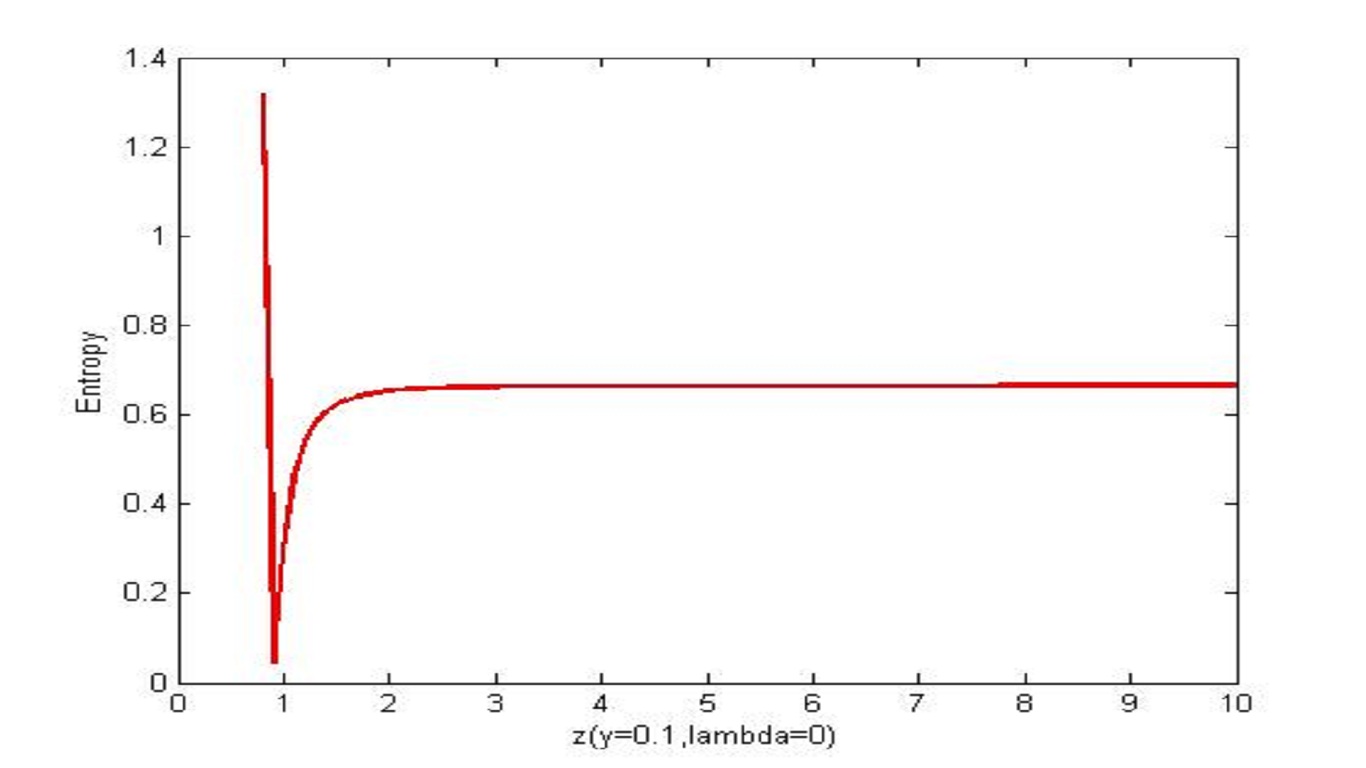}
 \caption{  the variation of $E(\protect\varrho )$ as a function of $z$ for fixed $y=0.1,\protect\lambda 
=0,y=0,\protect\alpha =1,q=0.01$.}
  \label{fig:PAVXRT01}
\end{center}
\end{figure}
Figure \ref{fig:PAVXTV02} shows the variation of $E(\varrho )$ as a function of $z$ and fixed $%
\lambda =0.01,y=0,\alpha =1,q=0.01$, this case is Schwarchild black hole in
noncommutative space-time.
\begin{figure}[htb!]
\begin{center}
  \includegraphics[width=6cm,height=5cm]{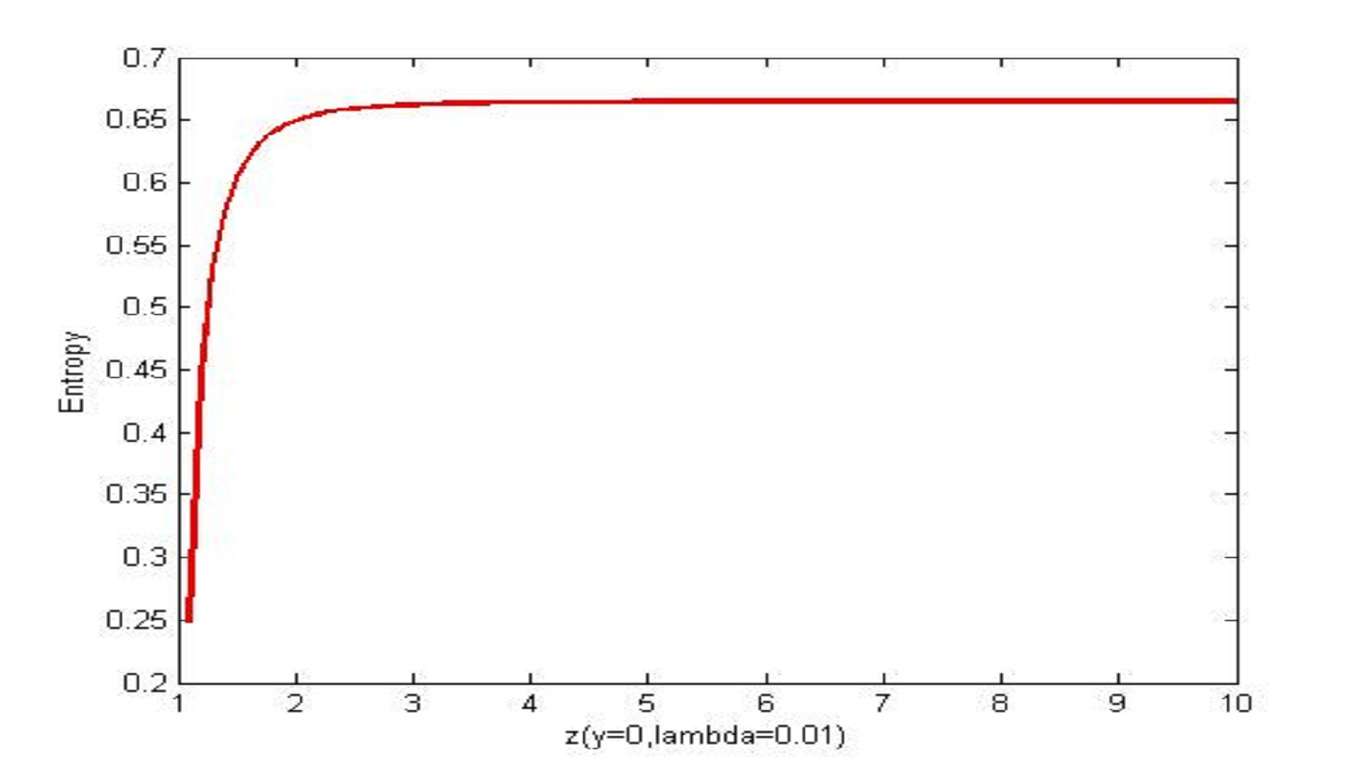}
 \caption{ the Entropy as a function of $z$ and fixed $\protect\lambda =0.01,y=0,\protect%
\alpha =1,q=0.01$.}
  \label{fig:PAVXTV02}
\end{center}
\end{figure}
Figure \ref{fig:PAVXV803} represents the variation of $E(\varrho )$ as a function of $z$ for
fixed $\lambda =0.1,y=2,\alpha =1,q=0.01,$case of Reissner Nordstrom Black
Hole in noncommutative space-time
\begin{figure}[htb!]
\begin{center}
  \includegraphics[width=6cm,height=5cm]{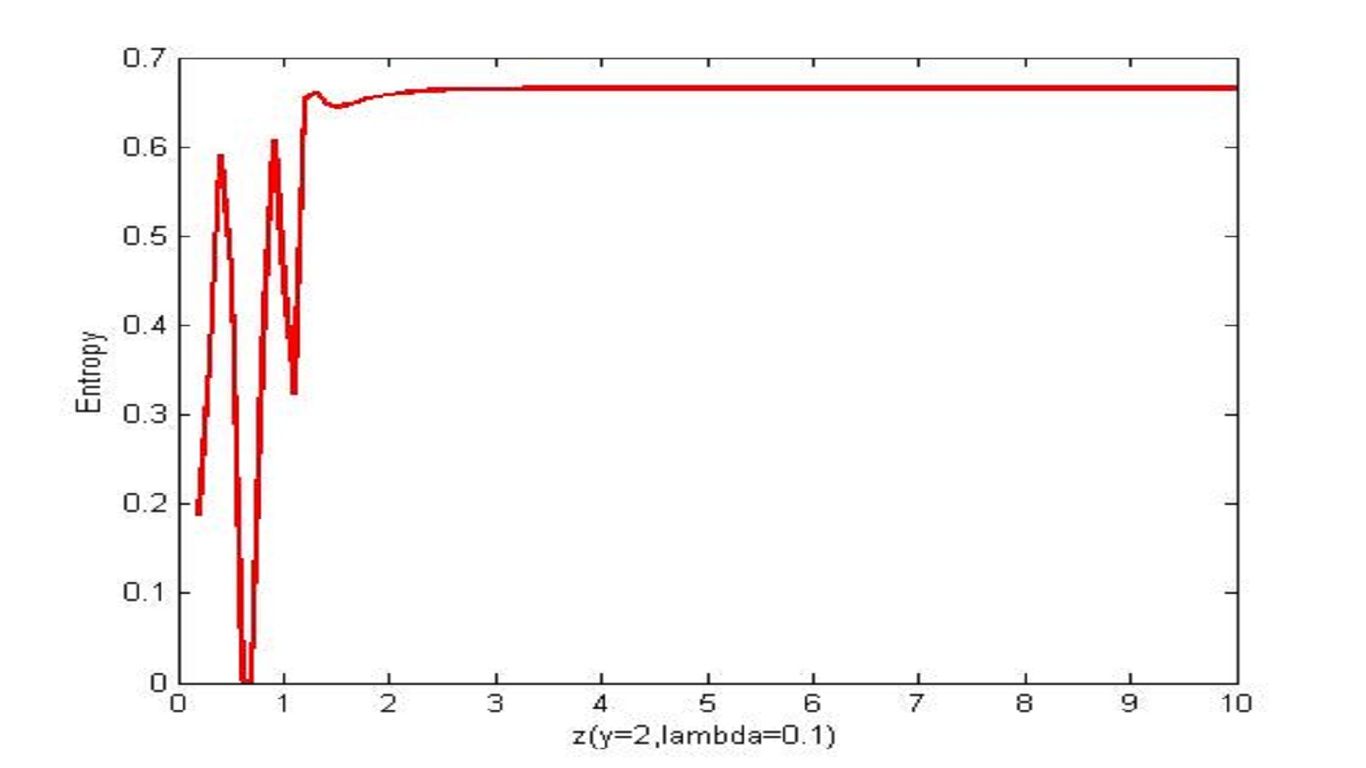}
 \caption{  the variation of $ E(\protect\varrho )$ as a function of $z$ for fixed $\protect\lambda =0.1,y=2,\protect\alpha =1,q=0.01$.}
  \label{fig:PAVXV803}
\end{center}
\end{figure}
Figure 12 shows the variation of the quantum entanglement entropy (EE) $E(\varrho )$
as a function of $z$ (or $r$). Notice that far from the oscillatory behavior
region, when $z$ (or $r$) increases, the GF $\widehat{g}_{00}$ decreases
until reaching a saturation value ($\approx 1$) where $E(\varrho )$ is
maximal, the oscillatory behavior disappears when we enter the stability
region where $E(\varrho )$ $\sim 0.67$. The explanation of the oscillatory
behavior is the same as for Figure2. The number of picks and minima depend
strongly on the values of the various parameters $\lambda ,y,\alpha $ and $q$%
. Concerning the non commutavity effect on the EE, it is clear that from eq(\ref{AI}) that for smaller values of $z$, as $\widehat{\eta }$ increases the
gravitational field $\widehat{g}_{00}$ becomes more important (increases)
and therefore $E(\varrho )$ decreases. For larger values of $z$, the effect
is almost negligeable since the terms $\sim \frac{1}{z^{4}},\frac{1}{z^{5}},%
\frac{1}{z^{6}}$ decreases faster than the commutative terms $\sim \frac{1}{z%
},$ Notice also that $y$ increases the GF increases ( the term $%
\widehat{\eta }^{2}\widehat{D}Q^{4}$  dominates at larger value of $Q$). Thus, the NC
effect on the EE becomes more imporatant for charged black hole than the
neutral ones ( if the charge $Q$ increases EE decreases ). Table \ref{tab:table1}
Sumarizes the effect og the black hole charge on the EE. It worth to mention that
in order to keep the perturbative expantion with respect to  ${\eta }^{2}$ reliable, one has to have
\begin{equation}
\left\vert \widehat{\eta }^{2}A_{1}\right\vert \prec \left\vert
A_{0}\right\vert 
\end{equation}%
where $A_{0}=\widehat{A}+\widehat{B}Q^{2}$ \ and $A_{1}=(\widehat{D}Q^{4}+%
\widehat{C}Q^{2}+\widehat{F}),$ this implies new constraints on the space
parameters $\lambda ,z,y,\alpha $
\begin{table}[htb!]
\begin{center}
\begin{tabular}{ |c|c|c|c|c| } 
\hline
$z$ & $2$ & $4$ & $5$ & $6$ \\  
\hline
$E(\varrho ),(y=0)$ & 0.64694 & 0.664 & 0.6644 & 0.6646 \\ 
\hline
$E(\varrho ),(y=10)$ & 0.6072 & 0.6567 & 0.6626 & 0.6641\\ 
\hline
\end{tabular}
\end{center}
\caption{\label{tab:table1} Illustrative values of EE as a function of z for y=0 and y=10}
\end{table}

\section{Conclusion}
Throughout this paper, we have studied in detail the singlet state of spin entanglement of two particles systems quantified by wootters concurrence and EE in a general static space-time. And applications, we have considered the Kerr and non commutative Reissner-Nordstrom space-time. In fact, in the first case we have studied the variation of the Wootters concurrence (WC) as a function of the various parameters such as q center of mass momentum of the wave packet, $\Sigma$ ( black hole rotation parameter), z (distance from the black hole). It turns out that the behavior of the WC depends strongly on those parameters(see figures 1,2,3,4,5 and 6). Regarding the second case (see figures 7,8,9,10,11,12) the variation of the quantum EE as a function of z, the NC parameter $\lambda $, the black hole charge y is discussed. We have noticed that the NC effect on the EE becomes more important in a charged black hole( more studies of triplet states and spin-momentum entanglement are under investigation)  
\renewcommand{\abstractname}{Acknowledgements}
\begin{abstract}
We are very grateful to the Algerian Minister of Higher Education and Scientific Research and DGRSDT for the financial support.
\end{abstract}
\clearpage

\bibliographystyle{plain} 
\bibliography{example}

\end{document}